\documentclass[10pt,aps,prr,final,twocolumn,amsmath,amssymb,superscriptaddress,longbibliography]{revtex4-1}
\usepackage[pdftex, colorlinks=true, linkcolor=blue, citecolor=blue, urlcolor=blue]{hyperref}
\usepackage{graphicx}
\usepackage{aurical}
\usepackage[T1]{fontenc}
\usepackage[version=3]{mhchem}
\usepackage{siunitx}
\usepackage{braket}
\graphicspath{{./figures/}}
\DeclareGraphicsExtensions{.png,.pdf}

\newcommand{\una}{Universit\'{e} de Nouakchott, Facult\'{e} des Sciences et Techniques, D\'{e}partement de Physique, Avenue du Roi Fai\c{c}al, 2373, Nouakchott, Mauritania}

\begin{document}

\title{Scaling Laws of Magnetically Driven High-order Harmonic Generation in Spin-Orbit Coupled Systems}

\author{O. Ly}
\email{ousmanebouneoumar@gmail.com}
\affiliation{\una}

\begin{abstract}
We investigate the scaling behavior of high harmonic generation (HHG) driven by magnetic dynamics in spin-orbit coupled systems. In contrast to optically driven HHG—where the harmonic cutoff scales as \(\omega^{-3}\) with the driving frequency \(\omega\)—our time-dependent quantum transport simulations reveal a qualitatively distinct scaling law for magnetically driven HHG in the presence of spin-orbit interaction: the harmonic cutoff \(n_{\mathrm{max}}\) scales as \(\omega^{-1}\). This fundamental difference arises from distinct excitation mechanisms—namely, spin-flip transitions driven by vectorial magnetic precession, as opposed to scalar electric fields. Furthermore, we demonstrate that the precession cone angle \(\theta\) serves as a crucial control parameter. Increasing \(\theta\) broadens the harmonic bandwidth, with peak emission achieved for nearly in-plane magnetic dynamics. Our findings establish magnetically driven HHG as a robust and tunable mechanism for nonlinear spin transport, governed by unique scaling laws with potential applications in ultrafast spintronic technologies.
\end{abstract}

\maketitle

\section{Introduction}

%The interplay between magnetization dynamics and spin transport lies at the core of modern spintronics. 
When a normal metal is placed adjacent to a ferromagnetic or antiferromagnetic material undergoing coherent magnetization precession, the conservation of angular momentum mandates the emission of a pure spin current into the nonmagnetic region \cite{Tserkovnyak2002, Tserkovnyak2002b, Cheng2014, Vaidya2020}. This process, known as \emph{spin pumping}, provides a robust, contactless means of generating spin currents through dynamic angular momentum exchange. It has been extensively studied and experimentally verified across a broad range of material systems \cite{Czeschka2011, Ando2011, Shiomi2014, Rojas2016, Song2016, Lesne2016, Song2017}.

Under steady-state precession at a driving frequency \(\omega\), the spin current emitted into the nonmagnetic region is time-periodic and oscillates at the same frequency. For a magnetization vector precessing with cone angle \(\theta\), the spin current takes the form:
\begin{equation}
\mathcal{J}(t) \propto \left( \sin{2\theta} \cos{\omega t}, \sin{2\theta} \sin{\omega t}, -2 \sin^2{\theta} \right),
\end{equation}
where the in-plane components oscillate sinusoidally, and the out-of-plane component remains constant. This regime corresponds to conventional spin pumping, dominated by harmonic content at the fundamental frequency \(\omega\).

Recent theoretical proposals suggest that coupling such magnetic dynamics to materials with strong spin-orbit coupling (SOC)—particularly Rashba systems—can dramatically enrich this behavior \cite{Ly2022}. In such systems, the interplay among precessing magnetic order, \(s\)-\(d\) exchange interaction, and SOC introduces strong nonlinearities into carrier dynamics. As a result, the system exhibits high harmonic generation (HHG): a nonlinear response in which spin and charge currents contain significant spectral weight at integer multiples of the driving frequency \(\omega\). This behavior goes beyond conventional spin pumping, yielding a rich and complex frequency spectrum.

This form of magnetically induced HHG represents a qualitatively new mechanism within the broader landscape of strong-field phenomena. 
{While HHG has been mainly explored in atomic gases since early nineties \cite{Krause1992, Huillier1993, Lewenstein1994, Popmintchev2012, Li2020, Zhang2007}, the effect has been further extended to solid state systems, including bulk crystals \cite{Ghimire2011} and Graphene \cite{Naotaka2017, Hafez2018}.  Later on it was studied in other carbon based structures such as $C_{60}$ \cite{Zhang2020} and Graphene nanobubbles \cite{Hadadi2023}. This, in addition to a variety of topological materials \cite{Jia2019, Li2022, Ono2024, Luka2024} and systems with strong SOC \cite{Lysne2020}. Furthermore, other effects such as strong electronic correlations \cite{Imai2020, Murakami2024} and altermagnetism \cite{Werner2024} have been shown to play an important role in enabling HHG in relevant condensed matter systems. In all these studies, the high harmonic response is driven by optical excitations. 
In contrast,} the mechanism studied here is driven by magnetization dynamics rather than optical stimuli. In both cases, harmonic generation emerges from strongly nonlinear processes shaped by system symmetries and coupling strengths. However, the magnetic route relies on time-dependent magnetic textures—specifically, the coherent precession of the magnetization vector—as the driving force. This introduces qualitatively different control parameters and offers a promising alternative pathway to generating nonlinear spin and charge responses, particularly in systems where direct optical excitation is impractical or spin-selective control is desired.

The generation and spectral structure of harmonics in magnetically driven Rashba systems are governed by several key parameters: the \(s\)-\(d\) exchange coupling strength \(J\), the Rashba SOC strength \(\alpha_{\mathrm{R}}\), the driving frequency \(\omega\), and the magnetization precession cone angle \(\theta\). In previous work \cite{Ly2022}, ultrahigh harmonic generation was predicted in the resonant regime where the Rashba splitting matches the magnetic gap, \( \alpha_{\rm R} k_{\mathrm{F}} = 2J \) ($k_{\mathrm{F}}$ being the Fermi wavelength), and the focus was placed on characterizing the extreme emission that emerges under this resonance condition. However, the influence of the precession frequency and cone angle on the harmonic cutoff was not systematically explored.

In this work, we shift our attention toward uncovering the scaling laws that govern the growth of the harmonic cutoff with respect to the magnetization dynamics---specifically, the precession cone angle \(\theta\) and the driving frequency \(\omega\). By doing so, we identify the key dynamical parameters that control the onset and extent of the ultrahigh harmonic regime. While strong SOC continues to play a crucial role by facilitating the nonlinear mixing of spin and orbital degrees of freedom, our analysis emphasizes how the geometric and temporal features of the magnetic drive shape the overall emission spectrum.

Although magnetically induced HHG has also been demonstrated in more complex systems such as non-collinear spin textures and topological materials~\cite{Ly2023}, our focus remains on homogeneous Rashba systems under uniform precession. This setting provides a minimal yet rich platform for identifying universal scaling trends, offering clear guidance for both theoretical modeling and experimental realization of magnetic HHG.
\begin{figure}
	\includegraphics[width=0.5\textwidth]{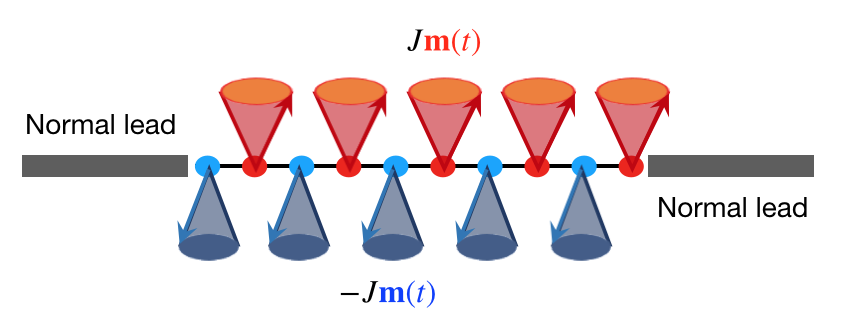}
	\caption{A sketch of the studied one dimensional antiferromagnetic system, hosting a Rashba SOC. The cones depicts the staggered  antiferromagnetic dynamics. The system is attached to two normal leads. Both magnetic moments oscillate with the same frequency but in opposite directions. 
    %This simple geometry is slightly different from the system investigated in \cite{Nikolic2022}, 
    This simple geometry is slightly different from the more general antiferromagnetic resonance setup,
    where the two moments precess with different cone angles. While, in the latter case both a staggered ($\mathbf{N}=(\mathbf{m}_a-\mathbf{m}_b)/2$) order and a ferromagnetic like  ($\mathbf{M}=(\mathbf{m}_a+\mathbf{m}_b)/2$) contribute to the pumped spin current. In the former case, only the staggered contribution $\mathbf{m}$ is expected. Here, $a$ and $b$ stand for the two sub-lattices of the antiferromagnet. In general, the pumped spin current in the absence of SOC is given as \cite{Cheng2014} $\mathcal{J}\propto \mathbf{N}\times\partial_t\mathbf{N}+\mathbf{M}\times\partial_t\mathbf{M}.$ In the actual case of a compensated antiferromagnet, only the first term survives. Once the SOC is turned on the dynamics becomes very non-linear leading to ultrahigh frequencies.
	}
	\label{fig:sketch}
\end{figure}
\begin{figure*}
    \includegraphics[width=\textwidth]{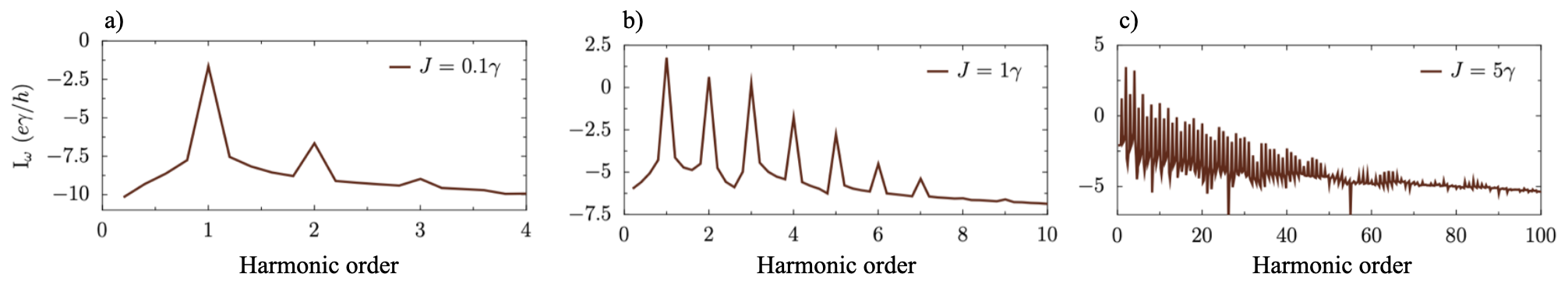}
	\caption{The HHG spectra at different values of the s-d exchange parameter are shown. Here, $J$ is varied from the low coupling case $J=0.1\gamma$ 
    %(corresponding to the studies in \cite{Nikolic2022, Manjarres2023}) 
    to the extremely nonlinear regime $J=5\gamma$. The parameters of the drive are chosen such that  $\hbar\omega=10^{-2}\gamma$ and $\theta=\pi/8$. The Fermi energy is set at $E_F=0$. Here, the current response is represented in a logarithmic scale.}
	\label{fig:js}
\end{figure*}

\section{Methodology}
\label{sec:methodology}

To investigate HHG driven by magnetic dynamics in spin-orbit coupled systems, we employ a time-dependent quantum transport framework built on a tight-binding model. Our analysis considers both one-dimensional (1D) and two-dimensional (2D) systems where a precessing magnetic order interacts with itinerant electrons subject to Rashba SOC. The systems are connected to normal metallic leads to allow for open-boundary, steady-state transport.

\subsection{Tight-Binding Model and Magnetic Dynamics}
The electronic structure is described using a nearest-neighbor tight-binding Hamiltonian on a square or linear lattice with lattice spacing $a$. The Hamiltonian includes kinetic hopping terms, Rashba spin-orbit interaction, and a local exchange coupling to a time-dependent magnetic texture:
\begin{equation}
\mathcal{H}(t) = \mathcal{H}_0(t) + \mathcal{H}_{\rm R},
\end{equation}
with
\begin{equation}
\mathcal{H}_0(t) = 
-\gamma \sum_{\langle \mathbf{r},\mathbf{r}' \rangle} c_{\mathbf{r}}^\dagger c_{\mathbf{r}'} 
+ J \sum_\mathbf{r} c_\mathbf{r}^\dagger (\mathbf{m}(\mathbf{r}, t) \cdot \boldsymbol{\sigma}) c_\mathbf{r},
\end{equation}
and
\begin{equation}
\label{eq:ham}
\mathcal{H}_{\rm R} =  \alpha_{\rm R} \sum_{ \mathbf{r} }i c_{x,y}^\dagger (\sigma_y c_{x+1,y} - \sigma_x c_{x,y+1}) + \rm{H.c},
\end{equation}
where \(\gamma\) is the hopping amplitude, \(\alpha_{\rm R}\) the Rashba SOC strength, \(J\) the $s$-$d$ exchange coupling, and \(\mathbf{m}(\mathbf{r}, t)\)  the unit vector representing the local magnetization direction at site \(\mathbf{r}=(x, y)\). The $2\times 2$ matrices \({\sigma_x}\) and \({\sigma_y}\) are the usual Pauli matrices. The additional term in Eq. (\ref{eq:ham}) stands for the hermitian conjugate of the first term. 
The antiferromagnetic dynamics is modeled as follows 
\begin{equation}
\mathbf{m}(\mathbf{r}, t) = (-1)^{x+y}(\sin\theta \cos \omega t, \sin\theta \sin \omega t, \cos\theta),
\end{equation}
leading to a staggered dynamics in the two-dimensional lattice. The parameters \(\theta\) and \(\omega\) depict respectively the precession angle and frequency of the vectorial magnetic drive. 
{The magnetic precession is assumed to result from the interaction between the magnetic moments and an external microwave field in the presence of an additional static field exerting a torque on the magnetization vector which then precess around the applied magnetic field. In the case of ferromagnetic systems the precession occurs in the GHz regime. In antiferromagnets the dynamics rather operate at frequencies in the THz range.}

\subsection{Time-Dependent Quantum Transport}
To simulate non-equilibrium transport, we compute the time-evolution of scattering states using Kwant and Tkwant packages \cite{kwant, Kloss2021}. The stationary scattering wave-functions $\Psi$ of the system are first computed at a given transport energy \(\varepsilon\), and then evolved in time by solving the time-dependent Schrödinger equation:
\begin{equation}
i\hbar \frac{d}{dt} \Psi(t) = \mathcal{H}(t) \Psi(t).
\end{equation}
{Defining \(\Psi_{lm\varepsilon}^{i}(t)\) as the time dependent $m^{th}$ component of the scattering wave function at site $i$, coming from lead $l$ with transport energy $\varepsilon$, the local current flowing between sites $i$ and $j$ at time $t$ can be obtained as :}
%The resulting time-dependent scattering wavefunctions \(\Psi_{lm\varepsilon}^{i}(t)\) allow us to compute the local currents at flowing between sites \(i\) and \(j\) as follows:
\begin{equation}
\label{eq:it}
{\rm I}_{ij}(t) = \sum_{lm} \int \frac{d\varepsilon}{\pi} \Im\left\{ \Psi_{lm\varepsilon}^{i \dagger}(t) \mathcal{H}_{ij}(t) \Psi_{lm\varepsilon}^{j}(t) \right\},
\end{equation}
where \(\mathcal{H}_{ij}(t)\) is the underlying Hamiltonian matrix element and \(\Im\) denotes the imaginary part. The total current flowing into a lead is obtained by summing over all interface bonds connecting the scattering region to the lead.

\subsection{Fourier Analysis and Harmonic Spectrum}

To extract the harmonic content, we perform a Fourier transform of the time-dependent current:
\begin{equation}
{\rm I}_\omega = \left| \int dt\, {\rm I}(t) e^{i\omega t} \right|,
\end{equation}
{where ${\rm I}(t)$ results from a sum of ${\rm I}_{ij}$ (Eq.~\ref{eq:it}) over the pairs of sites belonging to the  interface between the central system and the considered lead.}
The resulting spectrum reveals the distribution and cutoff of harmonics generated by the magnetic dynamics. We define the harmonic cutoff \(n_{\mathrm{max}}\) as the highest significant harmonic above background noise.

For robustness, we examine both single-energy (non-integrated) and full transport energy-integrated results. In all simulations, convergence with respect to energy integration, time discretization, and lead absorption was carefully verified. Furthermore, a special care was taken to suppress spurious reflections in the leads by tuning the maximum allowed reflection amplitude  to \(r_{max}=10^{-10}\). % and using sufficient absorbing layers within the metallic leads.

\section{Scaling up of the High Harmonic Emission}
%\section{Results}
\label{sec:scaling}
The main parameters that govern HHG are respectively: the s-d exchange coupling $J$, the opening of the dynamics $\theta$ and the driving frequency $\omega$. In this section, we examine the scaling up of HHG with respect to these parameters.  

\subsection{Dependence of HHG on the s-d exchange coupling $J$}
\label{sec:sd}
\begin{figure}%[htbp]
    \centering
	\includegraphics[width=0.5\textwidth]{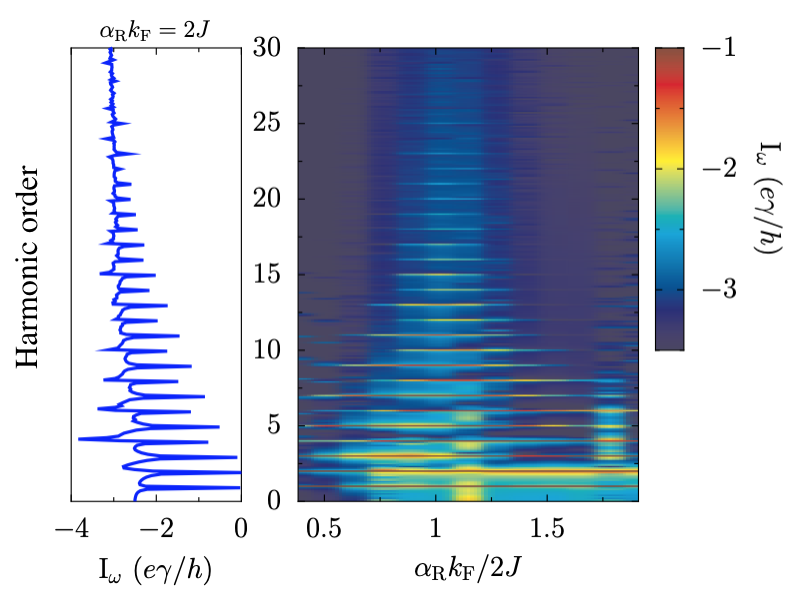}
	\caption{In the main panel, the Fourier transform of the charge current as a function of \(J\) is shown. Here, the calculations are performed at \(\alpha_{\rm R} a^{-1}= \gamma \). A resonance feature centered around \(\alpha_{\rm R} k_{\rm F} = 2J\) is observed. In the left panel, the Fourier spectrum at resonance is displayed. The driving frequency, the cone opening, and the Fermi energy are set to their values in Fig.~\ref{fig:js}. Here, the current response is represented in a logarithmic scale.}
	\label{fig:map}
\end{figure}

%To investigate the scaling laws underlying magnetically induced high harmonic generation (HHG), we first 
We consider a minimal model: a one-dimensional antiferromagnetic system with staggered magnetic order, driven at a frequency \(\hbar \omega = 10^{-2}\gamma\). The system includes Rashba SOC, a key ingredient for enabling high harmonic responses. The magnetic region is connected to two semi-infinite normal leads, as illustrated in Fig.~\ref{fig:sketch}.
We use Eq.~(\ref{eq:it}) to compute the charge current flowing into the right lead, with the Fermi energy fixed at \(E_{\rm F} = 0\). To probe how the \(s\)-\(d\) exchange coupling \(J\) influences the harmonic spectrum, we evaluate the current response for several values of \(J\).
In Fig.~\ref{fig:js}(a), we recover the low-coupling regime (\(J = 0.1\gamma\)), corresponding to the perturbative limit, which was the main focus of Ref.~\cite{Manjarres2023}.  
As shown in Fig.~\ref{fig:map}, \(\alpha_{\rm R} \) and \(J\) play similar roles in enabling high harmonic emission, with the overall strength primarily determined by their ratio. In this perturbative limit, only the first few harmonics appear prominently.
As the exchange coupling \(J\) increases, the harmonic spectrum becomes substantially broader, with harmonics extending up to nearly the 100th order. To access this highly nonlinear regime, we use a strong Rashba coupling comparable to the exchange parameter \(\alpha_{\rm R} a^{-1} = 6\gamma \). Nonetheless, enhanced harmonic emission can still be achieved at lower \(\alpha_{\rm R}\), provided \(J\) is comparable to or exceeds the hopping amplitude \(\gamma\).
Figure~\ref{fig:map} shows results for a reduced Rashba strength \(\alpha_{\rm R} k_{\rm {F}}= \gamma \) where \(J\) is varied. The emission peaks when the Rashba-induced splitting matches the antiferromagnetic exchange gap. This resonance condition, previously identified in Ref.~\cite{Ly2022}, and studied in \cite{Ly2025} from pure band dynamics perspective, occurs at \(\alpha_{\rm R} k_{\rm F} = 2J\) for the antiferromagnetic system. For ferromagnetic dynamics, the resonance shifts to \(\alpha_{\rm R} k_{\rm F} = J\), reflecting the halved band-gap in that case.
{We note that the coupling parameter $J$ used here falls well within the conventional range commonly employed in other theoretical studies. Earlier works~\cite{Xiao2008, Chen2009} have considered exchange couplings as large as \( J = 3.825\,\gamma \), which are comparable to the highest value \( J = 5\,\gamma \) adopted here. If we consider \( \gamma = 0.5 \, \rm{eV} \), the corresponding coupling constant would be  $J=2.5 \, \rm{eV}$, which is sufficiently high to place our calculations in the strongly nonlinear emission regime. 
}

Our results demonstrate that the spectral width of the emitted harmonics grows with increasing \(J\). However, the detailed structure of the spectrum does not vary monotonically with \(J\), due to the strong nonlinear dependence on both \(J\) and \(\alpha_{\rm R}\).
In both Fig.~\ref{fig:js} and Fig.~\ref{fig:map}, the harmonic structure reflects intrinsic features of the non-equilibrium dynamics. 

\subsection{Scaling up of HHG with respect to the driving frequency $\omega$}
\begin{figure*}[htbp]
	\includegraphics[width=\textwidth]{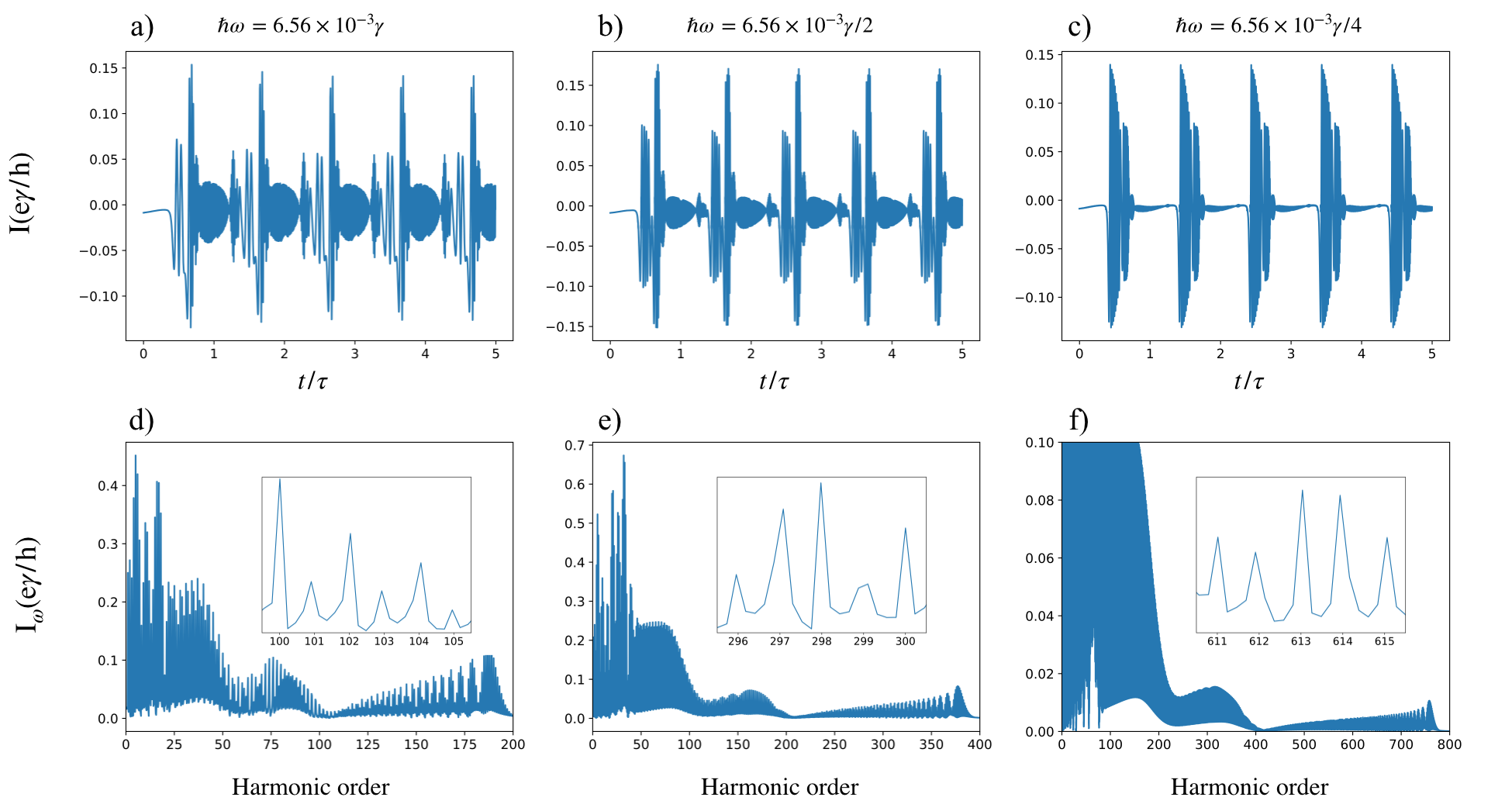}
	\caption{The scaling of HHG in terms of the driving frequency $\omega$ is shown. The upper row shows the time domain signals. In the lower row, the corresponding frequency domain responses are displayed. In the insets of d) - f), a zoom over few harmonics is shown. One observes clearly the increase of the HHG cutoff as $\omega$ is reduced. In panel f), the vertical axis is cut in order to zoom over the harmonics near the cutoff, whose amplitudes can be up to two orders of magnitude smaller than the intensity of the harmonics in the low harmonics region.}
	\label{fig:scale2d}
\end{figure*}
To study the scaling up of the magnetically driven HHG in terms of the driving frequency, 
%Here, 
we consider a compensated two-dimensional antiferromagnet coupled to a Rashba spin-orbit region, connected to two normal leads. %This differs from the one-dimensional case, where magnetism and SOC coexist in the same spatial region. 
{While in the previous subsection, the collective dynamics was studied, that is the contribution of all single energy levels, here we limit our attention to the single energy transport, which already captures the HHG features.}
Figure~\ref{fig:scale2d} shows the time-domain current signals and their Fourier transforms of the two dimensional antiferromagnetic system. To maximize harmonic emission, we tune the system to the resonant condition \(\alpha_{\rm{R}} k_{\rm F} = 2J\), aligning the Rashba splitting with the antiferromagnetic gap. At a single energy, the current exhibits up to 800 harmonics with substantial amplitude.
Even higher cutoffs, up to \(n_{\rm{max}} = 1200\), are achievable by increasing the cone angle \(\theta\) beyond \(30^\circ\). However, to highlight clear nonlinear scaling trends, we focus on a moderate value of \(\theta = 22.5^\circ\), which suppresses harmonics beyond \(n_{\rm{max}} = 800\). 
To study the growth of the harmonic cutoff with respect to the driving frequency $\omega$, we vary the latter and examine the resulting HHG signals. As shown in Fig.~\ref{fig:scale2d}(d–f), the number of harmonics increases significantly as \(\omega\) decreases. A similar trend holds for spin currents, though not shown here.

\begin{figure}
    %\centering
	\includegraphics[width=0.5\textwidth]{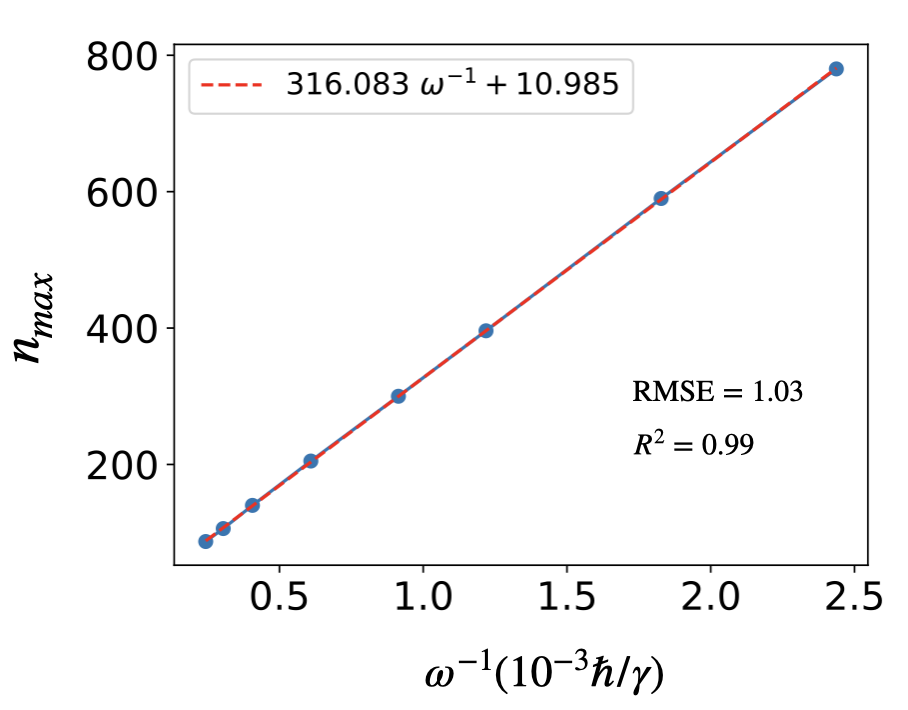}
	\caption{The scaling up of $n_{max}$ in terms of the driving frequency is shown. The sd exchange coupling and precession angle are set respectively at $J=\alpha_{\rm R}a^{-1}=5\gamma$ and $\theta=\pi/10$. 
    {To examine the fitting procedure used here, we define the root square mean error as ${\rm{RMSE}}^2=\langle(y_i-f_i)^2\rangle$, with $y$ and $f$ being respectively the original data and the fitting line. We also introduce the coefficient of determination $R^2$ given as $R^2=1-\sum_i(y_i-f_i)^2/\sum_i(y_i-\langle y\rangle)^2$. The value $R^2=0.99$ reflects an excellent linear fit, corresponding to an average deviation of a single harmonic (${\rm{RMSE}}=1.03$).}
    }
	\label{fig:omegas}
\end{figure}

At first glance, the number of harmonics \(n_{\rm{max}}\) appears to scale linearly with \(1/\omega\). {This scaling can intuitively be understood as a result of multiple spin-flip scattering events during a precession cycle. 
This phenomenological explanation suggests that when magnetization evolves much more slowly than the intrinsic spin-orbit timescale, more spin-flip events could occur per precession cycle, leading to a broad harmonic content. 
With sufficiently slow driving, cutoffs beyond \(n_{\rm{max}} = 1000\) are in fact attainable.}
Although such high harmonic orders may seem unusual for magnetic systems, they are consistent with trends in optically driven HHG, where cutoffs beyond \(n_{\rm{max}} = 5000\) have been reported \cite{Popmintchev2012}. The key determinant is the ratio of the driving frequency to the relevant microscopic scales: exchange interaction and SOC.
To quantify the dependence on \(\omega\), Fig.~\ref{fig:omegas} shows the maximum harmonic order observed \(n_{\rm{max}}\) versus the inverse frequency \(\omega^{-1}\). The relationship is clearly linear: \(n_{\rm{max}} \propto \omega^{-1}\). This contrasts sharply with laser-driven HHG, where the typical scaling is \(n_{\rm{max}} \propto \omega^{-3}\) \cite{Krause1992, Ghimire2019, Lysne2020}. The difference highlights the fundamentally distinct mechanisms in magnetic versus optical HHG.

\begin{figure*}
	\includegraphics[width=\textwidth]{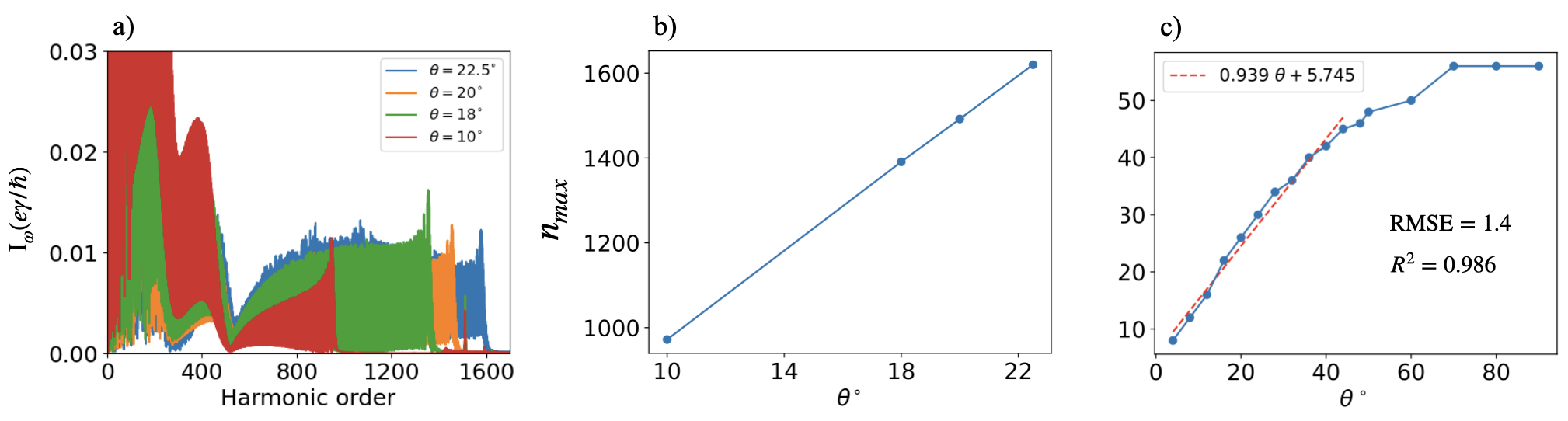}
	\caption{The evolution of the HHG bandwidth in terms of the precession angle is displayed. a) The Fourier spectra at different angles ranging from 10$^\circ$ to 22.5$^\circ$ are shown. Leading to an enhancement of the bandwidth by more than 600 harmonics. The driving frequency is taken such that  $\hbar\omega=0.0013\gamma$.  b) The quantitative scaling of $n_{max}$ with respect to $\theta$ is displayed. The data presented in panel a) are exploited. c) The scaling of $n_{\rm{max}}$ in terms of $\theta$ in the low emission regime is shown. The red dashed line corresponds to a fitting line of $n_{max}$ in the low angle branch. In c) a one dimensional antiferromagnet is considered (as in Fig.~\ref{fig:js} and Fig.~\ref{fig:map}). The parameters are chosen as $J=\alpha_{\rm R}a^{-1}=3\gamma$, $\omega=10^{-2}\gamma/\hbar$ and $E_{\rm F}=0$. Here, the current response is represented in linear scale. In panel c), the same fitting procedure as in Fig.~\ref{fig:omegas} is introduced.
    }
	\label{fig:angles}
\end{figure*}

\subsection{Scaling up of the Harmonic Cutoff with Respect to the Cone Opening of the Dynamics}
\label{sec:angles}
To gain a deeper understanding of HHG driven by magnetic dynamics, it is crucial to highlight a fundamental distinction from conventional laser-driven HHG. In magnetic systems, the driving field is not a scalar oscillation but a three-component vector field corresponding to the precession of local magnetic moments. This precessional motion traces a cone and its opening angle \( \theta \) plays a central role in shaping the carrier response.
Unlike laser-driven HHG, where the electric field is typically a scalar in nature, the magnetization vector in magnetic systems undergoes full three-dimensional motion. The angle \( \theta \) between the magnetization and the precession axis (typically aligned along the \( z \)-axis) introduces an additional degree of freedom that modulates the coupling between the magnetic drive and the spin-orbit-coupled carriers.
Although the driving frequency \( \omega \) primarily controls the global spectral bandwidth of harmonic generation, the amplitude and distribution of the emitted harmonics are strongly influenced by the cone angle \( \theta \).
Figure~\ref{fig:angles} illustrates the evolution of the HHG cutoff as a function of $\theta$. As shown, increasing \( \theta \) significantly broadens the harmonic bandwidth.
Notably, Fig.~\ref{fig:angles} a) reveals the emergence of harmonics around the 1500\textsuperscript{th} order. These features are attributed to transient dynamics near \( t = 0 \), when the magnetization rapidly departs from equilibrium.
The most favorable conditions for broadband harmonic emission arise in the fully in-plane precession limit (\( \theta = \pi/2 \)), where the magnetization lies entirely within the plane perpendicular to the precession axis. In this regime, coupling to conduction electrons is maximized, resulting in a broad and intense harmonic spectrum that extends deeply into the high-frequency regime.
Similar behavior is observed for large intermediate angles (e.g., \( \theta \approx \pi/3 \)), although the spectral intensity is slightly reduced compared to the fully in-plane configuration.
{Figure~\ref{fig:angles}(b) shows the dependence of the high-harmonic cutoff \( n_{{max}} \) on the cone angle \( \theta \). In the small-angle regime, \( n_{{max}} \) increases approximately linearly with \( \theta \), indicating the onset of the ultrahigh harmonic regime where large numbers of harmonics emerge. Minor fluctuations in \( n_{{max}} \) within this regime do not alter the overall linear scaling trend.
To quantify this behavior more precisely, we examine the low-emission regime (Fig.~\ref{fig:angles}(c)), where small deviations from linearity are more evident. This analysis confirms that the linear scaling of \( n_{{max}} \) with \( \theta \) holds over a broad range of precession angles, and appears to persist up to a saturation point beyond \( 50^\circ \). Although this linear dependence may seem counterintuitive, it parallels the well-known linear scaling of HHG cutoffs with driving field strength in optically driven systems~\cite{Ghimire2019}.
} {It is important to note that the results presented here are based on single-energy transport calculations. When integrated over the full energy spectrum, the ultrahigh harmonic response is expected to become even more pronounced.
These findings underscore the importance of magnetic drive geometry—particularly the precession cone angle—as a tunable parameter for controlling and enhancing nonlinear carrier responses. This opens promising avenues for engineering magnetically driven platforms capable of producing broadband HHG.
}

{
\section{Discussion and Conclusion}
In this work, we have investigated the scaling behavior of HHG induced by magnetization precession in spin-orbit coupled systems. Unlike conventional HHG---typically driven by intense optical fields and dominated by recollision dynamics of ionized carriers---our focus lies on a fundamentally distinct mechanism: the coherent, three-dimensional precession of a magnetic order parameter acting as a vectorial driving field. This alternative driving paradigm introduces rich geometric and dynamical degrees of freedom, offering new handles to control and enhance nonlinear responses in solid-state systems.

A central result of our study is the emergence of robust scaling laws governing the high-harmonic cutoff \( n_{\mathrm{max}} \), which denote the highest significant harmonic order observed in the spectrum. Specifically, we demonstrate that:
\(
n_{\mathrm{max}} \propto \omega^{-1},
\)
where \( \omega \) is the precession frequency of the magnetization. This inverse relationship implies that lower-frequency magnetic dynamics yield broader harmonic spectra, enabling access to higher harmonics without increasing the strength of the drive. This behavior stands in contrast to the cubic inverse scaling (\( n_{\mathrm{max}} \propto \omega^{-3} \)) commonly found in laser-driven HHG, where the cutoff is constrained by the ponderomotive energy and field recollision times \cite{Krause1992}. The comparatively mild scaling observed here reflects the fundamentally different physical origin of the harmonics---namely, spin dynamics and angular momentum exchange---rather than electron acceleration in real space.

In addition to frequency dependence, we identify a second crucial control parameter: the magnetization precession cone angle \( \theta \). For moderate angles, our simulations reveal a linear scaling of the harmonic cutoff with respect to the cone opening:
\(
n_{\mathrm{max}} \propto \theta.
\)
This relationship can be understood as a geometric enhancement: larger precession angles increase the transverse components of the magnetic field, thereby strengthening the time-dependent coupling between the local magnetization and spin-orbit coupled conduction electrons. While conventional ferromagnetic systems typically exhibit small precession cones (e.g., \( \theta \lesssim 20^\circ \)), more flexible platforms---{such as nano-oscillator devices \cite{Kiselev2003, Liu2012} are capable of supporting significantly larger cone angles, making them ideal candidates for observing the predicted scaling behavior and accessing the ultrahigh harmonic regime.

These findings point to the possibility of tailoring the HHG response through a combination of dynamical tuning and material design. Importantly, the conditions required for observing this ultrahigh harmonic regime are not merely theoretical constructs but are compatible with realistic material parameters. In particular, achieving \( J \sim \alpha_{\mathrm R} \) in the electronvolt range is feasible in materials such as bulk Rashba semiconductors~\cite{Eremeev2013, Sakano2013, Li2021} and topological insulators~\cite{Zhang2009, Liu2010}, where \( \alpha_{\mathrm R} a^{-1}\) can reach up to \( 4 \, \mathrm{eV} \). For instance, setting the hopping parameter to \( \gamma = 0.5 \, \mathrm{eV} \), our highest considered exchange value of \( J = 5\gamma \) corresponds to a realistic scale of \( 2.5 \, \mathrm{eV} \), well within the range observed in experiments.

In summary, our findings establish two key scaling laws for magnetization-driven HHG:
\(
n_{\mathrm{max}} \propto \omega^{-1}, \quad n_{\mathrm{max}} \propto \theta,
\)
providing both a dynamical and geometric route toward accessing ultrahigh harmonics in spin-orbit coupled materials. These scaling relations are not only theoretically transparent but also experimentally accessible, opening the door to compact, tunable, and material-integrated sources of high-frequency radiation.
}

\acknowledgments
We thank A. Manchon, X. Waintal and A. Abbout for useful discussions. 

\bibliography{refs}
\end{document}